# Effects of Topological Defect on the Energy Spectra and Thermo-magnetic Properties of $CO$ Diatomic Molecule.

## C. O. Edet and A. N. Ikot

Theoretical Physics Group, Department of Physics, University of Port Harcourt, Choba, Nigeria.

**ABSTRACT**

Confinement effects of Aharonov-Bohm (AB) flux and magnetic fields with topological defect on $CO$ diatomic molecule modeled by screened modified Kratzer potential is investigated in this paper. The all-encompassing effects of the fields and topological defect result in a strongly repulsive system. We discover that the collective effect of the fields and defect is intense than the lone and dual effect and consequently there is a substantial shift in the bound state energy of the system. We also find that to sustain a low-energy medium for the molecule modeled by SMKP, the topological defect and weak AB field are required, whereas the Magnetic field can be used as a control parameter or enhancer. The effects of the topological defect and magnetic and AB fields on the thermal and magnetic properties of the system are duly analyzed. We observe that the system tends to exhibit both a paramagnetic and diamagnetic behavior for weak and intense magnetic field respectively and some sort of saturation at large magnetic field. To further validate our findings, we map our result to 3D and a comparison of our results with what obtains in literature reveals an excellent agreement.

1. **INTRODUCTION**

The Schrödinger equation (SE) is a central equation in Quantum Mechanics (QM) which describes how the quantum state of a physical system changes with time. However, in practice, all the information for a quantum system is given by the solution of the SE [1-4]. The solutions to the SE yield the energy spectrum and wave function of the system under consideration. This can be obtained if we know the form of the core interaction potential. Many works concerning solutions of the Schrödinger equation with sundry potentials have been carried out extensively by researchers [5-10]. One of this potential amongst many others is the screened modified Kratzer potential (SMKP) [11] that was recently proposed by Edet et al [11]. The screened modified Kratzer potential is given by [11]

$$V(r) = D_e \left( \frac{qr - r_e e^{-\delta r}}{r} \right)^2 \qquad (1)$$

where $D_e$ is the dissociation energy, $r_e$ is the equilibrium bond length, $\delta$ is the screening parameter, $q$ is the control parameter whose choice is not a random one at all as will be shown later and the $r$ is the internuclear distance. In addition, we state here that this model (eq. (1)) is an improvement of the well-known modified Kratzer potential proposed by Babaei-Brojeny & Mokari [12], where we have introduced the exponential term (with the screening parameter) and the control parameter. The newly proposed model is a general case of the modified Kratzer which is obtained by setting $q=1$ and $\delta=0$, inversely quadratic Yukawa potential obtained by setting $q=0$ and many more models can be obtained from this model (1) with correct mapping. In a maiden report, Edet et al.,[11] solved the SE with the SMKP via the newly proposed Nikiforov-Uvarov-Functional Analysis (NUFA) method to obtain the energy spectrum and the corresponding wave function. Thermal properties of $CO$, $NO$ and $N_2$ diatomic molecules was obtained via the Euler–Maclaurin approach [11].

In fundamental quantum mechanics, the motion of a charged particle can be influenced by electromagnetic (EM) fields in regions in which the particle is rigorously confined [13,14]. In this region, if the EM field vanishes, the aftermath is a phenomenon known as the AB effect. The significance of the EM potential in quantum mechanics demonstrated by Aharonov and Bohm [15] in their seminal paper has initiated innovative considerations and revealed interesting quantum effects [15] associated with geometric phases. In the present day, geometric phases have been studied in condensed matter physics [16], holonomic quantum computation [17-20] and persistent currents in mesoscopic rings and quantum dots [21-30]. However, myriads of research articles considering the effects of magnetic and AB flux fields have been presented by several researchers. For instance, Falaye et al. [31] scrutinized the behaviour of the energy spectra of the hydrogen atom in a quantum plasma as it interacts with an electric field and exposed to linearly polarized intense laser field radiation. Eshghi and Mehraban [32] solved analytically the time-independent Dirac-Weyl equation for charge carriers with q-deformed pseudoscalar magnetic barrier (PMB) in graphene via the ansatz method. The authors also obtained a solution that explains the left propagating wave function for the calculation of the reflection and transmission coefficients using the Riemann's equation. It was also concluded that the Dirac-Weyl equation with PMB can help to comprehend the quantum behavior of the Dirac fermions. Eshghi and Mehraban [33] obtained analytically solutions for a graphene Dirac electron in magnetic fields with various q-parameters under an electrostatic potential. The Dirac–Weyl equation was solved with the Nikoforov–Uvarov (NU) and Frobenius methods. Thermal properties was obtained using the Hurwitz zeta function method for one of the states [33] and other numerous studies [34-40].

A pertinent and alluring problem in physics is to study the thermal and magnetic properties of quantum systems. Some researchers have carried investigations in this path. For example, Chargui and Dhahbi [41] investigated the statistical properties of a two-dimensional Dirac oscillator (2D DO) in the presence of a spin–orbit coupling, breaking its supersymmetry. The high temperature limit of different thermodynamical functions was probed analytically by utilizing the Euler–Maclaurin formula. Valentim et al. [42] studied a two-dimensional isotropic harmonic oscillator with a hard-wall confining potential in the form of a circular cavity defined by the radial coordinate $\rho_0$. The thermodynamics properties of the system was computed and the results compared to the well-studied free oscillator. Eshghi et al. [43] solved the SE with a position-dependent mass (PDM) for a charged particle with the superposition of the Hulthen plus Coulomb-like potential field under the influence of external magnetic and Aharonov–Bohm (AB) flux fields. The thermal quantities of the system was also considered and other many studies in this direction have been considered by several authors [44-50].

The equivalent of the EM AB effect is the background space-time of a cosmic string [51]. In this situation, the geometry is flat ubiquitously apart from the symmetry axis. Cosmic strings [52] and monopoles [53] are unusual topological defects [54] which were probably formed during phase transitions in the actual "early" history of the Universe.

Topology performs an imperative role in altering the physical properties of diverse quantum/physical systems and this has been a very valuable subject in different fields of physics, such as: gravitation and condensed matter physics. Topological defects materialize in gravitation as monopoles, strings and walls [55,56]. In condensed matter physics they are vortices in superconductors or superfluids [57], domain walls in magnetic materials [58], solitons in quasi-one-dimensional polymers [59] and dislocations or disclinations in disordered solids or liquid crystals [60]. The change in the topology of a medium introduced by a linear defect such as a disclination, dislocation or dispiration in an elastic media or a cosmic defect in space-time produces some effects on the physical properties of the medium [60].

A number of studies focusing on the influence of topological defects on semiconductors and quantum dots have attracted attention of researchers and this is quite evident in the literature [61-66]. Other works have dealt with topological defects in quantum rings [67, 68], Soheibi et al[69] studied electrons subject to the deformed Kratzer potential with screw dislocation [69]. Filgueiras et [70] investigated the effect of screw dislocation on the eigenvalues and eigenfunction of an electron confined in a 2D pseudoharmonic quantum dot under the influence of an external magnetic field inside a dot and AB field inside a pseudodot. In a recent work, Bakke and Furtado [71] considered an elastic medium with a disclination and studied the effect of topological defect on the interaction of a spinless electron with radial electric fields.

Based on the information gathered, it is thus, very expedient to study the influence of disclination on particle dynamics in the presence of AB and magnetic fields [66, 72]. However, to the best of our knowledge, no study in literature has scrutinized the effects of toplogical defects on $CO$ diatomic molecule. Therefore, the major goal of this paper is in two-fold, first we study the energy shift related with a non-relativistic quantum particle interacting with screened Modified Kratzer potential in the spacetimes generated by a cosmic string under the combined effects of the magnetic and AB fields for $CO$ diatomic molecule. More so, we analyse the effects of the topological defect on the thermal and magnetic properties of $CO$ diatomic molecule.

This paper is organized as follows. In section 2 we present the theory and calculations. In section 3, we evaluate the thermal and magnetic properties of the screened Modified Kratzer potential under the magnetic and AB fields with topological defect. In section 4, we discuss the effects of the topological defect on the energy spectra and thermal and magnetic properties of $CO$ diatomic molecule placed in the gravitational field of a cosmic string under the magnetic and AB fields. Finally, in section 5, we draw some conclusions.

## 2. THEORY AND SOLUTIONS

Let us consider that the diatomic molecule is modeled by screened modified Kratzer potential, under the all-inclusive effect of the AB and magnetic fields. Let us assume that there exist a disclination or topological defect in this region. The disclination is described by the line element [56, 66, 71]

$$ds^2 = dr^2 + \alpha^2 r^2 d\varphi^2 + dz^2 \tag{2}$$

where $0 < \alpha < 1$ is the parameter associated with the deficit of angle. The parameter $\alpha$ is related to the linear mass density $\tilde{\mu}$ of the string via $\alpha = 1 - 4\tilde{\mu}$. Notice that the azimuthal angle is defined in the range $0 \leq \varphi \leq 2\pi$ [66].

The Hamiltonian operator of a charged particle and subjected to move in the screened modified Kratzer potential under the combined impact of AB flux field and influence of external uniform magnetic field with topological defect can be written in cylindrical coordinates. Thus, the Schrödinger equation for this consideration is written as follows;

$$\left[ \frac{1}{2\mu}\left( i\hbar \vec{\nabla}_\alpha - \frac{e}{c}\vec{A} \right)^2 + \frac{D_e}{\alpha}\left( \frac{qr - r_e e^{-\delta r}}{r} \right)^2 \right] \psi(r,\varphi) = E_{nm} \psi(r,\varphi), \tag{3}$$

where $E_{nm}$ denotes the energy level, $\mu$ is the effective mass of the system, the vector potential can be written as a sum of two terms $\vec{A} = \vec{A}_1 + \vec{A}_2$ having the azimuthal components [43] $\vec{A}_1 = \frac{|\vec{B}|e^{-\delta r}}{\alpha(1-e^{-\delta r})}\hat{\varphi}$ and $\vec{A}_2 = \frac{\phi_{AB}}{2\pi r}\hat{\varphi}$, $\vec{B} = B\hat{z}$ is the applied external uniform magnetic field with $\vec{\nabla} \times \vec{A}_1 = |\vec{B}|$, $A_2$ represents the additional magnetic flux $\phi_{AB}$ created by a solenoid with $\vec{\nabla} \cdot \vec{A}_2 = 0$, [41]. The Del and Laplacian operators are used as in Refs. [71, 73].

Several forms of the vector potential exist, depending on the area of interest one can adopt a suitable form for a study. For instance Eshghi and Mehraban [32] solved the Dirac Weyl equation with various vector potentials. Eshghi and Mehraban [33] obtained analytically solutions for a graphene with various forms vector potential. However, when solving exactly solvable models, the vector potential of the form: $A = \frac{|\vec{B}|r}{2}$ for the magnetic field have been adopted by many authors [35,38]. Falaye et al. [31] adopted the vector potential of the form: $A(t) = \left(\frac{\varepsilon_0}{\omega}\right)\cos(\omega t)$ to study Hydrogen atom in a Laser plasma environment. Ferkous and Bounames[34] solved the 2D Pauli equation with Hulth´en potential for spin-1/2 particle in the presence of Aharonov–Bohm (AB) field with the vector potential of the form; $A = -\frac{\alpha}{r}u_\theta$. Eshghi et al. [43] solved the SE with a position-dependent mass (PDM) for a charged particle with the superposition of the Hulthen plus Coulomb-like potential field under the influence of external magnetic and Aharonov–Bohm (AB) flux fields. The authors used the vector potential of the simple form as [31, 41];

$$\vec{A} = \left(0, \frac{|\vec{B}|e^{-\delta r}}{\alpha(1-e^{-\delta r})} + \frac{\phi_{AB}}{2\pi r}, 0\right). \tag{4}$$

We are motivated to adopt the vector potential of the form in Eq. (4) above since our model is of similar form to that of Eshghi et al. [43]. Our choice of the vector potential of the form is inspired by the work of Eshghi et al. [43]. Moreover, we introduced the parameter $\alpha$ to show the influence of the topological defect in the medium, this is done following the work of Bakke and Furtado [71].

Let us take a wave function in cylindrical coordinates as $\psi(r,\varphi) = \frac{1}{\sqrt{2\pi r}} e^{im\varphi} R_{nm}(r)$, where $m$ denotes the magnetic quantum number. Inserting this wave function and the vector potential into Eq. (3) we arrive at the following radial second-order differential equation:

$$R''_{nm}(r) + \frac{2\mu}{\hbar^2}\left[E_{nm} - V_{eff}(r)\right] R_{nm}(r) = 0 \tag{5}$$

where $V_{eff}(r)$ is the effective potential defined as follows;

$$V_{eff}(r) = \frac{D_e}{\alpha}\left(\frac{qr - r_e e^{-\delta r}}{r}\right)^2 + \hbar\omega_c\left(\frac{m}{\alpha^2} + \frac{\xi}{\alpha}\right)\frac{e^{-\delta r}}{(1-e^{-\delta r})r} + \left(\frac{\mu\omega_c^2}{2}\right)\frac{e^{-2\delta r}}{(1-e^{-\delta r})^2} + \frac{\hbar^2}{2\mu}\left[\frac{\left(\frac{m}{\alpha^2} + \xi\right)^2 - \frac{1}{4}}{r^2}\right] \tag{6}$$

where $\xi = \frac{\phi_{AB}}{\phi_0}$ is an integer with the flux quantum $\phi_0 = \frac{hc}{e}$ and $\omega_c = \frac{e|\vec{B}|}{\mu c}$ denotes the cyclotron frequency.

Eq. (5) is not exactly solvable due to the presence of centrifugal term. Therefore, we employ the Greene and Aldrich approximation scheme [74] to overcome the centrifugal term. This approximation is given as:

$$\frac{1}{r^2} = \frac{\delta^2}{(1-e^{-\delta r})^2} \tag{7}$$

We point out here that this approximation is only valid for small values of the screening parameter $\delta$. Inserting Eqs. (7) into Eq. (6) and introducing a new variable $s = e^{-\delta r}$ allows us to obtain

$$\frac{d^2 R_{nm}(s)}{ds^2} + \frac{1}{s}\frac{dR_{nm}(s)}{ds} + \frac{1}{s^2(1-s)^2}\left[-\left(\varepsilon_{nm} + \rho_0 + \rho_1 + \Omega_1\right)s^2 + \left(2\varepsilon_{nm} + \rho_1 - \Omega_0\right)s - \left(\varepsilon_{nm} + \gamma\right)\right]R_{nm}(s) = 0 \quad (8)$$

For Mathematical simplicity, let's introduce the following dimensionless notations;

$$A = \frac{q^2 D_e}{\alpha}, B = \frac{D_e r_e^2}{\alpha}, C = \frac{2q D_e r_e}{\alpha}, -\varepsilon_{nm} = \frac{2\mu(E_{nm} - A)}{\hbar^2 \delta^2}, \rho_0 = \frac{2\mu B}{\hbar^2}, \rho_1 = \frac{2\mu C}{\hbar^2 \delta},$$

$$\Omega_0 = \frac{2\mu\omega_c}{\hbar\delta}\left(\frac{m}{\alpha^2} + \frac{\xi}{\alpha}\right), \Omega_1 = \frac{\mu^2 \omega_c^2}{\hbar^2 \delta^2} \text{ and } \gamma = \left(\frac{m}{\alpha} + \xi\right)^2 - \frac{1}{4} \quad (9)$$

In order to solve eq. (8), we have transformed the differential equation (8) into a form solvable by a standard mathematical technique. Hence, we take the radial wave function of the form

$$R_{nm}(s) = s^\lambda (1-s)^\sigma f_{nm}(s) \quad (10)$$

where

$$\lambda = \sqrt{\varepsilon_{nm} + \gamma} \quad (11a)$$

$$\sigma = \frac{1}{2} + \sqrt{\frac{1}{4} + \rho_0 + \Omega_1 + \Omega_0 + \gamma} \quad (11b)$$

On substitution of Eq. (10) into Eq. (8), we obtain the following hypergeometric equation:

$$f''_{nm}(s) = \frac{\left((2\lambda + 2\sigma + 1)s - (2\lambda + 1)\right)}{s(1-s)} f'_{nm}(s) + \frac{\left((\lambda+\sigma)^2 - \left(\sqrt{\varepsilon_{nm} + \rho_0 + \rho_1 + \Omega_1}\right)^2\right)}{s(1-s)} f_{nm}(s) = 0 \quad (12)$$

Equation (13) is a more suitable second order homogeneous linear differential equation, the solution of which can be achieved by using the well-known asymptotic iteration method [75]. The systematic procedure of the asymptotic iteration method begins now by rewriting equation (12) in the following form [75];

$$f''_{nm}(s) - \lambda_0(s) f'_{nm}(s) - s_0(s) f_{nm}(s) = 0 \quad (13)$$

where

$$\lambda_0(s) = \frac{\left((2\lambda + 2\sigma + 1)s - (2\lambda + 1)\right)}{s(1-s)} \text{ and } s_0(s) = \frac{\left((\lambda+\sigma)^2 - \left(\sqrt{\varepsilon_{nm} + \rho_0 + \rho_1 + \Omega_1}\right)^2\right)}{s(1-s)} \quad (14)$$

The primes of the function $f_{nm}(s)$ in equation (13) indicates the derivatives with respect to $s$. The asymptotic feature of the method for sufficiently large $k$ is given as [75, 76]

$$\frac{s_k(s)}{\lambda_k(s)} = \frac{s_{k-1}(s)}{\lambda_{k-1}(s)} = \alpha(s) \quad (15)$$

where

$$\lambda_k(s) = \lambda'_{k-1}(s) + s_{k-1}(s) + \lambda_0(s)\lambda_{k-1}(s),$$
$$s_k(s) = s'_{k-1}(s) + s_0(s)\lambda_{k-1}(s), \tag{16}$$

Equation (16) is referred to as the recurrence relations [76, 77]. In accordance with asymptotic iteration method [75, 77], the equation we seek can be obtained from the roots of the following equation [75]:

$$\delta_k(s) = \begin{vmatrix} \lambda_k(s) & s_k(s) \\ \lambda_{k+1}(s) & s_{k+1}(s) \end{vmatrix} = 0, \quad k = 1, 2, 3... \tag{17}$$

With the aid of the quantisation condition given in Eq. (17), we arrive at the following eigenvalues expressions

$$\delta_0(s) = \begin{vmatrix} \lambda_0(s) & s_0(s) \\ \lambda_1(s) & s_1(s) \end{vmatrix} = 0 \Leftrightarrow \lambda_0 = -0 - \sigma \pm \sqrt{\varepsilon_{nm} + \rho_0 + \rho_1 + \Omega_1}$$

$$\delta_1(s) = \begin{vmatrix} \lambda_1(s) & s_1(s) \\ \lambda_2(s) & s_2(s) \end{vmatrix} = 0 \Leftrightarrow \lambda_1 = -1 - \sigma \pm \sqrt{\varepsilon_{nm} + \rho_0 + \rho_1 + \Omega_1} \tag{18}$$

$$\delta_2(s) = \begin{vmatrix} \lambda_2(s) & s_2(s) \\ \lambda_3(s) & s_3(s) \end{vmatrix} = 0 \Leftrightarrow \lambda_2 = -2 - \sigma \pm \sqrt{\varepsilon_{nm} + \rho_0 + \rho_1 + \Omega_1}$$

...etc.

By finding the nth term of the above progression, we obtain;

$$\lambda_n = -n - \sigma \pm \sqrt{\varepsilon_{nm} + \rho_0 + \rho_1 + \Omega_1} \tag{19}$$

By carrying out some simple algebraic manipulations, and using the expressions in eq. (9), we obtain the energy of the screened modified Kratzer potential under influence of magnetic and AB fields with topological defects as follows;

$$E_{nm} = \frac{q^2 D_e}{\alpha} + \frac{\hbar^2 \delta^2}{2\mu}\left(\left(\frac{m}{\alpha} + \xi\right)^2 - \frac{1}{4}\right) - \frac{\hbar^2 \delta^2}{8\mu}\left[\frac{Q_1 - (n+Q_2)^2}{(n+Q_2)}\right]^2 \tag{20}$$

where

$$Q_1 = \frac{2\mu D_e r_e^2}{\hbar^2 \alpha} + \frac{4\mu q D_e r_e}{\hbar^2 \delta \alpha} + \frac{\mu^2 \omega_c^2}{\hbar^2 \delta^2} - \left(\left(\frac{m}{\alpha} + \xi\right)^2 - \frac{1}{4}\right)$$

$$Q_2 = \frac{1}{2} + \sqrt{\frac{2\mu D_e r_e^2}{\hbar^2 \alpha} + \frac{\mu^2 \omega_c^2}{\hbar^2 \delta^2} + \frac{2\mu \omega_c}{\hbar \delta}\left(\frac{m}{\alpha^2} + \frac{\xi}{\alpha}\right) + \left(\frac{m}{\alpha} + \xi\right)^2} \tag{21}$$

For completeness sake, we move to find the wave function of the system. Let us now calculate the wave function of this system. In general, the differential equation we wish to solve should be transformed to a form that is convenient for applying AIM [75, 76]:

$$y''(x) = 2\left(\frac{ax^{N+1}}{1-bx^{N+2}} - \frac{M+1}{x}\right)y'(x) - \frac{Wx^N}{1-bx^{N+2}}y(x) \tag{22}$$

where $a, b$ and $M$ are constants

The exact solutions for eq. (22) is given by

$$y(x) = (-1)^n C(N+2)^n (\sigma)_n {}_2F_1\left(-n, t+n; \Omega; bx^{N+2}\right) \tag{23}$$

where,

$$(\Omega)_n = \frac{\Gamma(\Omega+n)}{\Gamma(\Omega)}, \Omega = \frac{2M+N+3}{N+2}, t = \frac{(2M+1)b+2a}{(N+2)b} \tag{24}$$

By comparing equation (12) with (22), we can deduce that

$$M = \lambda - \frac{1}{2}, \ t = 2(\lambda+\sigma), a = \sigma, \Omega = 2\lambda+1, \ b = 1, N = -1, \ (\Omega)_n = \frac{\Gamma(2\lambda+1+n)}{\Gamma(2\lambda+1)} \tag{25}$$

and

$$f_{nm}(s) = (-1)^n C_2 \frac{\Gamma(2\lambda+1+n)}{\Gamma(2\lambda+1)} {}_2F_1\left(-n, 2(\lambda+\sigma)+n; 2\lambda+1; s\right) \tag{26}$$

It is therefore straight forward to deduce that the corresponding unnormalized wave function is obtain as;

$$R_{nm}(s) = (-1)^n N_{nm} \frac{\Gamma(2\lambda+1+n)}{\Gamma(2\lambda+1)} s^\lambda (1-s)^\sigma {}_2F_1\left(-n, 2(\lambda+\sigma)+n; 2\lambda+1; s\right) \tag{27}$$

The full wave function is written as follows;

$$\psi(r,\varphi) = \frac{1}{\sqrt{2\pi r}} e^{im\varphi} (-1)^n N_{nm} \frac{\Gamma(2\lambda+1+n)}{\Gamma(2\lambda+1)} \left(e^{-2\delta r}\right)^\lambda \left(1-e^{-2\delta r}\right)^\sigma {}_2F_1\left(-n, 2(\lambda+\sigma)+n; 2\lambda+1; e^{-2\delta r}\right)$$

The three dimensional non-relativistic energy solutions are obtained by setting $m = \ell + \frac{1}{2}$, in Eq. (21) to obtain;

$$E_{n\ell} = q^2 D_e + \frac{\hbar^2 \delta^2 \ell(\ell+1)}{2\mu} - \frac{\hbar^2 \delta^2}{8\mu} \left[\frac{\frac{2\mu D_e r_e^2}{\hbar^2} + \frac{4\mu q D_e r_e}{\hbar^2 \delta} - \ell(\ell+1) - \left(n+\frac{1}{2}+\sqrt{\frac{1}{4}+\frac{2\mu D_e r_e^2}{\hbar^2}+\ell(\ell+1)}\right)^2}{\left(n+\frac{1}{2}+\sqrt{\frac{1}{4}+\frac{2\mu D_e r_e^2}{\hbar^2}+\ell(\ell+1)}\right)}\right]^2 \tag{28}$$

where $\ell$ is the rotational quantum number

## 3. Thermal and Magnetic properties of screened modified Kratzer potential (SMKP)

It is established extensively in literature and in basic text [48, 50] that all thermodynamic properties can be obtained from the partition function of the system. This means that the successful evaluation of the partition function of the system is the commencing point to evaluate all other thermal functions of the system under consideration. The vibrational partition function can be computed by straightforward summation over all possible vibrational energy levels accessible to the system. Given the energy spectrum (37), the partition function $Z(\beta)$ of the screened modified Kratzer potential at finite temperature $T$ is obtained with the Boltzmann factor as [50];

$$Z(\beta) = \sum_{n=0}^{n_{max}} e^{-\beta E_n} \tag{29}$$

with $\beta = \dfrac{1}{kT}$ and with $k$ is the Boltzmann constant.

Substituting eq. (21) in (29), we have;

$$Z(\beta) = \sum_{n=0}^{n_{max}} e^{-\beta\left(\frac{q^2 D_e}{\alpha} + \frac{\hbar^2 \delta^2}{2\mu}\left(\left(\frac{m}{\alpha}+\xi\right)^2 - \frac{1}{4}\right) - \frac{\hbar^2 \delta^2}{8\mu}\left(\frac{Q_1-(n+Q_2)^2}{(n+Q_2)}\right)^2\right)} \tag{30}$$

where $n$ is the vibrational quantum number, $n = 0, 1, 2, 3 ... n_{max}$ , $n_{max}$ denotes the upper bound vibration quantum number. The maximum value $n_{max}$ can be obtained by setting $dE_n/dn = 0$ ,

$$n_{max} = -\left(\frac{1}{2} + \sqrt{\frac{2\mu D_e r_e^2}{\hbar^2 \alpha} + \frac{\mu^2 \omega_c^2}{\hbar^2 \delta^2} + \frac{2\mu \omega_c}{\hbar \delta}\left(\frac{m}{\alpha^2}+\frac{\xi}{\alpha}\right)+\left(\frac{m}{\alpha}+\xi\right)^2}\right) \pm \sqrt{Q_1} \tag{31}$$

For a finite summation with the maximum value $n_{max}$, the Poisson summation formula can be written as [48, 50];

$$\sum_{n=0}^{n_{max}} f(n) = \frac{1}{2}\left[f(0) - f(n_{max}+1)\right] + \sum_{M=-\infty}^{\infty} \int_0^{n_{max}+1} f(x) e^{-i2\pi M x} dx \tag{32}$$

Under the lowest-order approximation, neglecting the quantum corrections which include the terms with $M = 0$ , we write summation formula (32) as follows:

$$\sum_{n=0}^{n_{max}} f(n) = \frac{1}{2}\left[f(0) - f(n_{max}+1)\right] + \int_0^{n_{max}+1} f(x) dx . \tag{33}$$

Applying the above to equation (30) yields the following expression,

$$Z(\beta) = \frac{1}{2}\left[e^{-\beta(Q_0 - \Lambda p_1^2)} - e^{-\beta(Q_0 - \Lambda p_2^2)}\right] + \int_0^{n_{max}+1} e^{-\beta\left(Q_0 - \Lambda\left(\frac{Q_1}{(x+Q_2)} - (x+Q_2)\right)^2\right)} dx \qquad (34)$$

where we have defined the following for mathematical simplicity

$$\Lambda = \frac{\hbar^2 \delta^2}{8\mu}, \quad Q_0 = \frac{q^2 D_e}{\alpha} + \frac{\hbar^2 \delta^2}{2\mu}\left(\left(\frac{m}{\alpha} + \xi\right)^2 - \frac{1}{4}\right)$$

$$p_1 = \frac{Q_1}{Q_2} - Q_2 \qquad (35)$$

$$p_2 = \frac{Q_1}{n_{max} + 1 + Q_2} - (n_{max} + 1 + Q_2)$$

If we set $y = \frac{Q_1}{(x+Q_2)} - (x+Q_2)$, we can rewrite the above integral in eq.(34) as follows;

$$\int_0^{n_{max}+1} e^{-\beta\left(Q_0 - \Lambda\left(\frac{Q_1}{(x+Q_2)} - (x+Q_2)\right)^2\right)} dx = \frac{1}{2} e^{-\beta Q_0} \int_{p_1}^{p_2} e^{\beta \Lambda y^2} \left(\frac{y}{\sqrt{y^2 + 4Q_1}} - 1\right) dy \qquad (36)$$

$$= \frac{1}{2} e^{-\beta Q_0} \left(\begin{array}{c} \frac{\sqrt{\pi}\left(Erfi\left[p_1\sqrt{\beta}\sqrt{\Lambda}\right] - Erfi\left[p_2\sqrt{\beta}\sqrt{\Lambda}\right]\right)}{2\sqrt{\beta}\sqrt{\Lambda}} \\ - \frac{e^{-4Q_1\beta\Lambda}\sqrt{\pi}\left(Erfi\left[\sqrt{p_1^2 + 4Q_1}\sqrt{\beta}\sqrt{\Lambda}\right] - Erfi\left[\sqrt{p_2^2 + 4Q_1}\sqrt{\beta}\sqrt{\Lambda}\right]\right)}{2\sqrt{\beta}\sqrt{\Lambda}} \end{array}\right)$$

where Erfi denotes the imaginary error function, which is defined explicitly in ref. [44, 45].

Substituting equation (36) into (34), we obtain the following vibrational partition function of the screened modified Kratzer potential (SMKP) in magnetic and AB fields with topological defects as follows;

$$Z(\beta) = \frac{1}{2} e^{-\beta Q_0} \left[\begin{array}{c} e^{\beta\Lambda p_1^2} - e^{\beta\Lambda p_2^2} + \frac{\sqrt{\pi}\left(Erfi\left[p_1\sqrt{\beta}\sqrt{\Lambda}\right] - Erfi\left[p_2\sqrt{\beta}\sqrt{\Lambda}\right]\right)}{2\sqrt{\beta}\sqrt{\Lambda}} \\ - \frac{e^{-4Q_1\beta\Lambda}\sqrt{\pi}\left(Erfi\left[\sqrt{p_1^2 + 4Q_1}\sqrt{\beta}\sqrt{\Lambda}\right] - Erfi\left[\sqrt{p_2^2 + 4Q_1}\sqrt{\beta}\sqrt{\Lambda}\right]\right)}{2\sqrt{\beta}\sqrt{\Lambda}} \end{array}\right] \qquad (37)$$

This expression represents the classical partition function. The reason is that equation (37) does not contain quantum corrections.

In what follows, all thermodynamic and magnetic properties of the Screened modified Kratzer potential in the presence of the AB and magnetic fields with topological defect, such as the free energy, mean energy, the entropy, specific heat, magnetization, magnetic susceptibility and the persistent current, can be obtained from the partition function (37), $Z(\beta)$. These thermodynamic functions for the diatomic molecules system can be calculated from the following expressions[50, 79];

$$F(\beta) = -\frac{1}{\beta} \ln Z(\beta),$$

$$U(\beta) = -\frac{d \ln Z(\beta)}{d\beta},$$

$$S(\beta) = \ln Z(\beta) - \beta \frac{d \ln Z(\beta)}{d\beta},$$

$$C(\beta) = \beta^2 \frac{d^2 \ln Z(\beta)}{d\beta^2}, \qquad (38)$$

$$M(\beta) = \frac{1}{\beta}\left(\frac{1}{Z(\beta)}\right)\left(\frac{\partial}{\partial \vec{B}} Z(\beta)\right),$$

$$\chi_m(\beta) = \frac{\partial M(\beta)}{\partial \vec{B}},$$

$$I(\beta) = -\frac{e}{hc}\frac{\partial F(\beta)}{\partial m}$$

## 4. Applications

In this section, we apply our results obtained in the previous sections to study the *CO* diatomic molecules in the presence of magnetic and AB fields with topological defect. The *CO* diatomic molecule is chosen because of it wide application and studies by several authors. For instance, Jia et al [49] presented a study vibrational partition function for *CO* diatomic molecule with the improved Tietz potential energy model. Jia et al. [80] presented a categorical representation of molar entropy for gaseous substances with *CO* diatomic molecule inclusive. Jiang et al [81] presented a universal efficient closed-form representation for the molar enthalpy of *CO* diatomic molecule. We point out here that the authors of refs [49, 80, 81], studied the *CO* diatomic molecule in the absence of external fields and topological defect.

Moreover, As a result of engaging novel electrochemical methods, *CO* can be transformed into a great number of carbon fuels and other commodity chemicals [82-84]. The experimental values of molecular constants for lowest (i.e. ground) electronic state of the CO molecule are taken from literature [85]:

$D_e = 87471.43 \left(cm^{-1}\right)$, $r_e = 1.1282 \overset{o}{A}$, $\mu = 6.860586 amu$ and $\alpha = 2.29940 \overset{o}{A}^{-1}$. We used the following conversions; $\hbar c = 1973.269$ eV $\overset{0}{A}$ and $1 amu = 931.5 \times 10^6 eV \left(\overset{0}{A}\right)^{-1}$ for all computations[50].

The potential under study is well-behaved for both small and large inter-nuclear separations. This molecular potential model provides a realistic description of the rotating diatomic molecules and also could be useful in discussing long amplitude vibrations in large molecules. We analyse (graphically) the energy levels, magnetic and thermal properties of rotational $CO$ diatomic molecules in two dimension space when applying the magnetic and AB fields. More so, for validity we will also in 3D space without topological defect, magnetic and AB field compare the numerical energy obtained via AIM with the NUFA method.

Table 1 presents eigenvalues for the Screened Kratzer potential (SMKP) for $CO$ diatomic molecule under the influence and the absence of external fields (the magnetic field and AB flux field) and topological defect in $eV$ and in low vibrational $n$ and rotational $m$. From the table, in the absence of external fields and defect (i.e., when $|\vec{B}| = \xi = 0$ and $\alpha = 1$), the spacing between the energy levels of the effective potential is narrow and increases with increasing $n$. We notice that there exists degeneracy among some states $(n, m)$ [for instance: (0, 0), (0, 1), and (0,-1)]; (1, 0), (1, 1) and (1,-1); (2, 0), (2, 1), and (2,-1); and (3, 0), (3, 1), and (3,-1)], but application of the magnetic field strength not only increases the energy levels of the effective potential and spacings between states but also transforms the degeneracies to pseudo-degeneracies. Moreover, the pseudo-degeneracies among the states are also removed and the energy values shift up.

By subjecting the system to only the AB flux field, the energy values are reduced and degeneracies are removed, whereas the pseudo-degeneracies among the states are not affected. The energy levels become more positive and the system becomes strongly positive as the quantum number $n$ increases for fixed $m$.

When only the topological defect is present, the degeneracies and the pseudo-degeneracies are not completely altered and the repulsiveness of the total interaction potential increases. The topological defect also shifts the energy level upward.

The combined effect of the magnetic and AB fields, in the absence of the topological defect is stronger than the individual effect and absence of the fields. We notice that there is a substantial shift in the energy levels. When only the magnetic and topological defect are present. The effect is also stronger and there is a great shift in the energy levels but lesser when compared to the presence of the fields only. When only the AB field and topological defect are present. The effect is also stronger and there is a great shift in the energy levels but lesser than the previous two cases.

The complete effects indicate that the system is strongly repulsive. Also, the combined effect (triad) of the fields and topological defect is stronger than the individual effects and dual effects and consequently, there is a noteworthy shift in the bound state energy of the system.

In Fig. 1 of the present study we plot the Effective potential energy for $CO$ diatomic molecule with $(m=0)$ levels for (a) Effective potential energy against internuclear distance for various values of the topological defect $(\alpha)$. As shown on the plot, we observe that by increasing the topological defect and keeping all fields constant leads to a corresponding decrease in the effective potential function. Thus, the potential energy becomes more repulsive. Fig.1(b) Effective potential energy against internuclear distance for Various values of $|\vec{B}|$. We see that increasing the strength of the magnetic field increases does not really show an effect but influences the attractiveness of the effective potential. Fig. 1(c) Effective potential energy against internuclear distance for various values of $\xi$. We see that by setting effects of AB field to be constant and then varying the magnetic field up to 30T has little or no significant effect on the effective model.

In Fig. 2 we show the variation of the energy values of the SMKP for $CO$ diatomic molecule as a function of the magnetic field with varying defect. The effect of the AB flux field on the $CO$ diatomic molecule is stronger than that of the topological defect. As shown in Fig. 2, we see that as the magnetic field increases, the energy increases. The effect of the topological defect is great when small. For instance, when $\alpha=0.4$ is greater than when $\alpha=0.8$ $and$ 1 respectively. This is the same with the inset. But to emphasize the strength of the AB field, we observe that when the AB field is say $\xi=10$ is smaller compared to the $\xi=100$ in the same magnetic field interval of $0<|\vec{B}|<20T$, we notice a huge shift in the energy level and a decline in the energy at say $|\vec{B}|\approx 5T$.

In Fig. 3 shows the variation of energy values for the $CO$ diatomic molecule in SMKP under the influence of the magnetic field and AB flux fields with topological defect as a function of topological defect. We see clearly that the energy reduces with increasing $\alpha$ and AB field for $|\vec{B}|=10T$ but when one pays close attention to the inset where $|\vec{B}|=100T$, an energy shift is slightly noticed but not as higher as the AB field case.

In Fig. 4, we show plots of the Partition function as a function of $\beta$, topological defect $(\alpha)$, AB flux, $\xi$ and magnetic fields, $|\vec{B}|$. Fig 4 (a) clearly shows that the partition decreases with increasing $\beta$ for different values of $\alpha$. The partition increases for a fixed value of the topological parameter $(\alpha)$. We also see that the partition function is higher in the absence of the topological defect $(\alpha=1)$. Fig. 4 (b & c) displays the partition function as a function of $\beta$ for different values of $\xi$ and $|\vec{B}|$. It is seen that the partition function decreases with increasing $\beta$ in both cases. and $\xi$. We observe here that for fixed values of the magnetic and AB fields, the partition function decreases as fields increases. As a confirmation of our observation in Fig. 4(a), Fig. 4(d) shows that the partition increases as the topological defect increases but decreases with increasing $\beta$. Fig. 4(e) shows the plot of partition function as a function of $|\vec{B}|$ with different values of $\beta$. In this figure, we see that in the region $0 < |\vec{B}| < 10T$, the partition function first decreases and then rises beyond this region as the magnetic field becomes intense up $|\vec{B}| = 100T$. This peak is observed in the interval $10T < |\vec{B}| < 100T$. In Fig. 4 (f), we show the plot of the partition function as a function of $\xi$ different values of $\beta$. The partition function decreases with increasing AB field.

In Fig. 5, we show plots of the Helmholtz Free Energy as a function of $\beta$, topological defect $(\alpha)$, AB flux, $\xi$ and magnetic fields, $|\vec{B}|$. Fig. 5(a) clearly shows that the Free energy increases monotonically as $\beta$ increases for different values of $\alpha$. The free energy increases with increasing $\beta$. At $\alpha = 0.4$, the free energy is higsher than in the absence $(\alpha = 1)$. In Figs. 5(b&c), the free energy increases monotonically with increasing $\beta$ with the free energy in each case being higher at $\xi = 30$ and $|\vec{B}| = 10T$. The free energy is plotted as a function of $\alpha$ different values of $\beta$ in Fig. 5(d). The free energy decreases as $\alpha$ increases. In Fig. 5 (e), Free energy as a function of $|\vec{B}|$ different values of $\beta$. The free energy again decreases with increasing magnetic field. Fig. 5(f) shows a plot of free energy as a function of $\xi$ different values of $\beta$. The free energy increases linearly with rising AB field.

In Fig.6, we show plots of the Entropy as a function of $\beta$, topological defect $(\alpha)$, AB flux, $\xi$ and magnetic fields, $|\vec{B}|$. Fig. 6 (a) displays the Entropy as a function of $\beta$ for different values of $\alpha$. The entropy decays as $\beta$ increases. The entropy is higher for $\alpha = 1$. Figs 6 (b &c) entropy is plotted as a function of $\beta$ for different values of $\xi$ and $|\vec{B}|$. The entropy of the system increases with increasing $\beta$. The entropy is noted to be higher for $\xi = 30$ and $|\vec{B}| = 10T$. Figure 6 (d) displays the entropy of the system as a function of $\alpha$ for different values of $\beta$. The entropy of the system increases as $\alpha$ increases. Fig. 6 (e) shows a plot of entropy as a function of $|\vec{B}|$ different values of $\beta$. We observe that the entropy increases as the magnetic field decreases. Fig. 6 (f) shows a plot of Entropy as a function of $\xi$ different values of $\beta$. The entropy of the system reduces as the AB field increases.

In Fig.7, we show plots of the Mean or average energy as a function of $\beta$, topological defect $(\alpha)$, AB flux, $\xi$ and magnetic fields, $|\vec{B}|$. Fig. 7(**a**) demonstrates the plot of mean energy as a function $\beta$ for different values of $\alpha$. The mean energy decreases as $\beta$ increases. The topological defect is seen to have a great Impact as the mean energy is seen to be higher at $\alpha = 0.4$. Fig. 7 (b&c) shows plots of mean energy as a function of $\beta$ for different values of $\xi$ and $|\vec{B}|(T)$. The mean energy in this cases reduces as $\beta$ increasing. The mean energy is said to be higher for $\xi = 30$ and $|\vec{B}| = 10T$. Fig. 7 (d) shows the plot of mean energy as a function of $\alpha$ with different values of $\beta$. The mean energy decreases with increasing $\alpha$. Fig. 7 (e) shows the plot of mean energy as a function of $|\vec{B}|$ alongside different values of $\beta$. The mean energy increases with rising magnetic field, $|\vec{B}|(T)$. Fig. 7 (f) shows the plot of mean energy as a function of $\xi$ different values of $\beta$. The mean energy increases with rising AB field.

An analysis of the specific heat capacity of the system is extensively carried out in fig. (8). The variation pattern of heat capacity versus $\beta$ for $CO$ diatomic molecule is scrutinized. In Fig.8, we show plots of the Specific Heat Capacity as a function of $\beta$, topological defect $(\alpha)$, AB flux, $\xi$ and magnetic fields, $|\vec{B}|$. Fig. 8 (a) shows the specific heat capacity versus $\beta$ for different values of $\alpha$. The specific heat capacity increases with increasing $\beta$. The effect is stronger at $\alpha = 0.4$.

Also, figs. 8 (b & c) shows the specific heat capacity of SMKP as a function of $\beta$ for different values of $\xi$ and $|\vec{B}|(T)$ in that order. We observe here that the specific in these situation follows a similar pattern. The specific heat capacity initially upsurges with the rising $\beta$ up until it arrives at a maximum and then drops. Fig. 8 (d) shows the plot of specific heat capacity versus $\alpha$ different values of $\beta$. Again, the specific heat capacity at first rises with the rising $\beta$ up until it arrives at a maximum and then drops. Fig. 8 (e) shows the plot of specific heat capacity versus $|\vec{B}|$ different values of $\beta$. The specific heat capacity decreases as the magnetic field increases. Fig. 8 (f) shows the plot of specific heat capacity versus $\xi$ for different values of $\beta$. The specific heat increases as $\beta$ increases.

In Fig.9, we show plots of the Magnetization as a function of $\beta$, topological defect $(\alpha)$, AB flux, $\xi$ and magnetic fields, $|\vec{B}|$. Figs. 9(a-c) shows the plot of Magnetization as a function of $\beta$ for different values of $\alpha$, $\xi$ and $\vec{B}$. It is noted that the magnetization decreases with rising $\beta$. Here we notice high magnetization at $\alpha = 1, \xi = 10$ and $|\vec{B}| = 10T$ in that order. In fig. 9 (d) magnetization is plotted as a function of $\alpha$ different values of $\beta$. We observe here that the magnetization increases monotonically with increasing $\alpha$. Fig. 9 (e) shows the magnetization as a function of $|\vec{B}|$ different values of $\beta$. The magnetization at first declines with the rising $\beta$ up until it arrives at a minimum and then rises again. Fig. 9 (f) shows the magnetization as a function of $\xi$ different values of $\beta$. The magnetization increases with increasing AB field.

Fig. 10 show plots of the Magnetic Susceptibility as a function of $\beta$, topological defect $(\alpha)$, AB flux, $\xi$ and magnetic fields, $|\vec{B}|$. Fig 10 (a) shows the plot of Magnetic Susceptibility as a function of $\beta$ for different values of $\alpha$. We observe that the magnetic susceptibility peaks up as $\beta$ rises and the $\chi_m(\beta)$ decreases. $\chi_m(\beta)$ is also higher in the absence of the topological defect $(\alpha = 1)$. A similar trend is exhibited in Fig 10 (b) although $\chi_m(\beta)$ is higher when the AB field, $\xi = 30$. In the two cases discussed, the system is paramagnetic. An alike behavior is also displayed in Fig. 10 (c), in this case $\chi_m(\beta)$ is high at $|\vec{B}| = 10T$ and the system tends to exhibit both a paramagnetic behavior and diamagnetic behavior when $|\vec{B}| = 4T$ and $|\vec{B}| = 6T$. Fig. 10 (d) shows the plot of $\chi_m(\beta)$ as a function of $\alpha$ for different values of $\beta$. $\chi_m(\beta)$ at first declines with the rising $\beta$ up until it arrives at a minimum and then rises again.

In Fig. 10 (e) we $\chi_m(\beta)$ as a function of $|\vec{B}|$ different values of $\beta$. The plot reveals that as $|\vec{B}|$ increases, $\chi_m(\beta)$ increases and exhibits some sort of saturation at large $|\vec{B}|$ [86, 87]. This observation comes out to be a general one whenever we monitor the variation of $\chi_m(\beta)$ against $|\vec{B}|$ under different conditions. The plot also shows paramagnetic behavior of the system over an interval of $|\vec{B}|$. We observe a crossing behavior in the close neighborhood of $|\vec{B}| \sim 4T$. The crossing behavior may perhaps have its origin in the changing effective confinement of the system, coupled with the consequent modification in the extent of $CO$ molecule repulsive interaction. Fig. 10 (f) shows the plot of $\chi_m(\beta)$ versus $\xi$ different values of $\beta$. $\chi_m(\beta)$ at first rises with the rising of $\beta$ up until it arrives at a maximum and then drops.

We state here that the varying values/ranges of $|\vec{B}|$ (when the system changes from paramagnetism to diamagnetism and vice versa) depend on the particular values of $\chi_m(\beta)$ at which the observations are made.

In Fig.11, we show plots of the persistent current as a function of $\beta$, topological defect $(\alpha)$, AB flux, $\xi$ and magnetic fields, $|\vec{B}|$. Figs. 11 (a-c) shows the plots of persistent current as a function of $\beta$ for different values of $\alpha$, $\xi$ and $|\vec{B}|$. We observe that the persistent current decreases with rising $\beta$. Here we notice high magnetization at $\alpha = 1, \xi = 30$ and $|\vec{B}| = 10T$ in that order. In fig. 11 (d) persistent current is plotted as a function of $\alpha$ for different values of $\beta$. We observe here that the persistent current increases monotonically with increasing $\alpha$. Fig. 11 **(e)** shows the persistent current as a function of $|\vec{B}|$ for different values of $\beta$. The persistent current at first declines with the rising $\beta$ up until it arrives at a minimum and then rises again. Fig. 11 **(f)** shows the plot of persistent current as a function of $\xi$ different values of $\beta$. The persistent current increases with increasing AB field.

## 5. CONCLUSION

In this paper, we have studied the effects of the topological defect and magnetic and AB flux field on the energy spectra, thermal and magnetic properties of the screened modified Kratzer potential for $CO$ diatomic molecule. The complete influences show that the system is strongly repulsive while the localizations of quantum levels change and the eigenvalues decrease. Also, as we have established, the all-inclusive effect of the topological defect and fields is stronger than individual and dual effects and consequently there is a significant shift in the bound state energy of the system. We found that to maintain a low energy for the $CO$ diatomic molecule, the topological defect and AB field are required, whereas the magnetic field can be used as a controller or an enhancer. The effects of the topological defect and magnetic and AB fields on the thermal and magnetic properties of the system are duly analyzed. We observe that the system tends to exhibit both a paramagnetic and diamagnetic behavior for weak and intense magnetic field respectively and some sort of saturation at large magnetic field. To further validate our findings, we map our result to 3D and a comparison of our results with what obtains in literature reveals an excellent agreement. The present calculation scheme can be further used to study the thermal and magnetic properties of triatomic substance [88-90]. More so, thermal properties of $NO$, $N_2$, etc. diatomic molecules can also be studied [91-95]

## ACKNOWLEDGMENTS

The authors dedicates this work to (Dr Akpan Ndem Ikot) on his birthday anniversary. We also thank the anonymous referees for the careful reading and the suggestions that improved the paper.

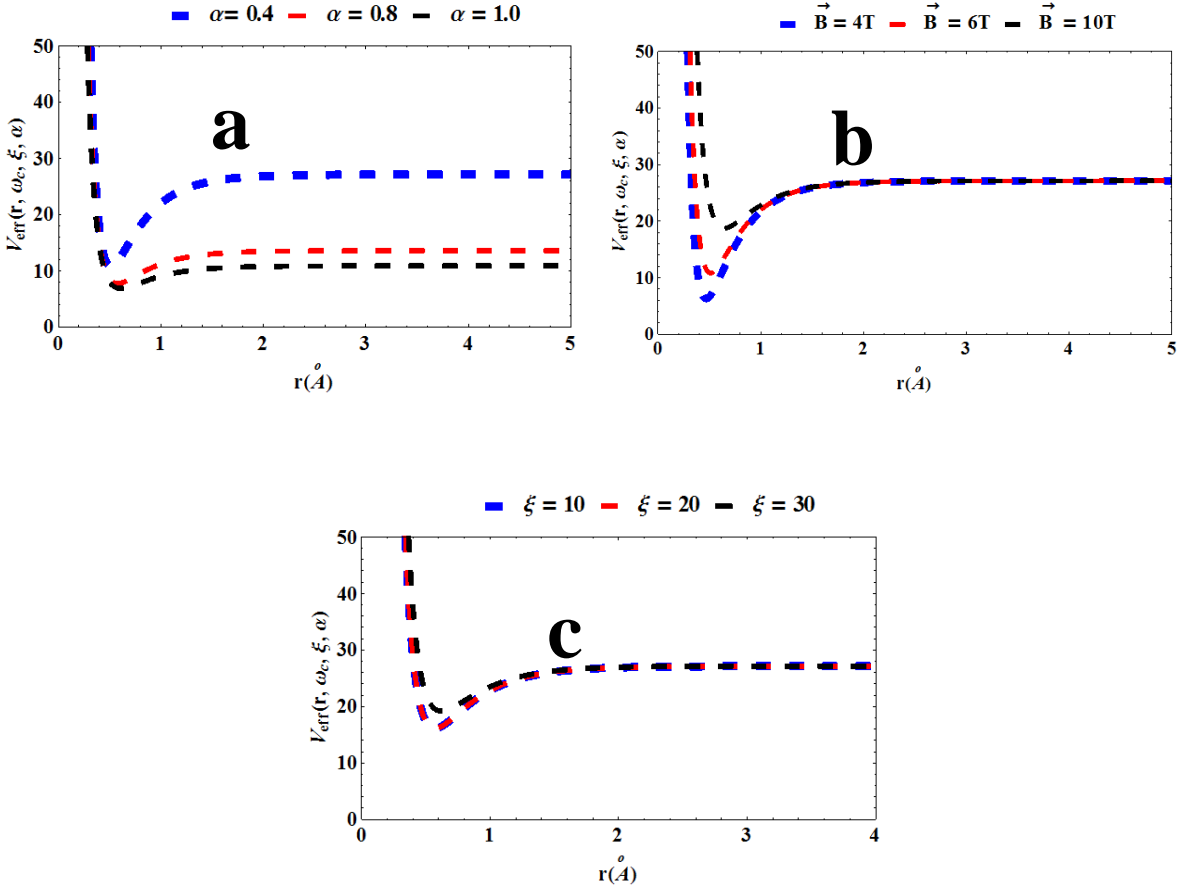

**Figure 1**: The Effective potential energy to $CO$ diatomic molecule with $(m=0)$ levels for **(a)** various values of the topological defect $(\alpha)$ with $|\vec{B}|=6T$ and $\xi=6$. Increasing the topological defect and keeping all fields constant leads to a corresponding decrease in the effective potential function. Thus, the potential energy becomes more repulsive. **(b)** Various values of $|\vec{B}|$ with $\xi=6$ and $\alpha=0.4$. **(c)** Various values of $\xi$ with $|\vec{B}|=6T$ and $\alpha=0.4$. Increasing the strength of the AB field increases does not really show an effect but influences the attractiveness of the effective potential. Setting effects of AB field to be constant and then varying the magnetic field up to 30T has little or no significant effect on the effective model.

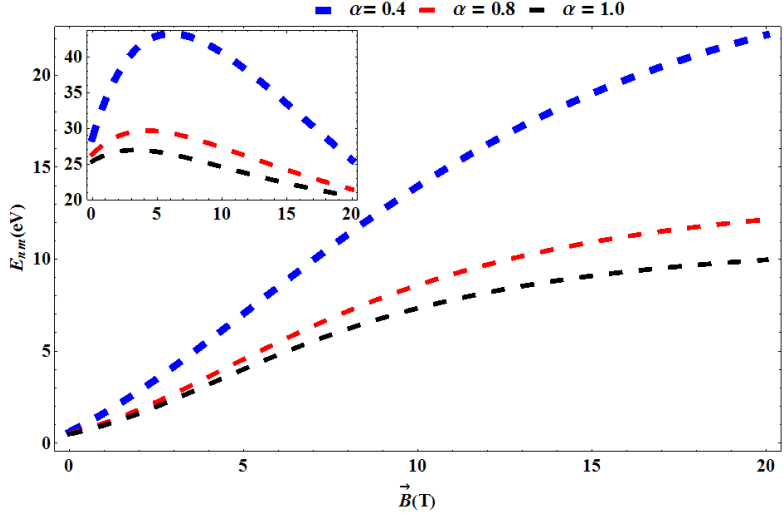

**Figure. 2**: Variation of energy values for the $CO$ diatomic molecule in SMKP under the influence of the magnetic field and AB flux fields with topological defect. Using the fitting parameters $m=n=0$ and $\xi=10$ as a function of external magnetic field with various $\alpha$; the inset is for $\xi=100$.

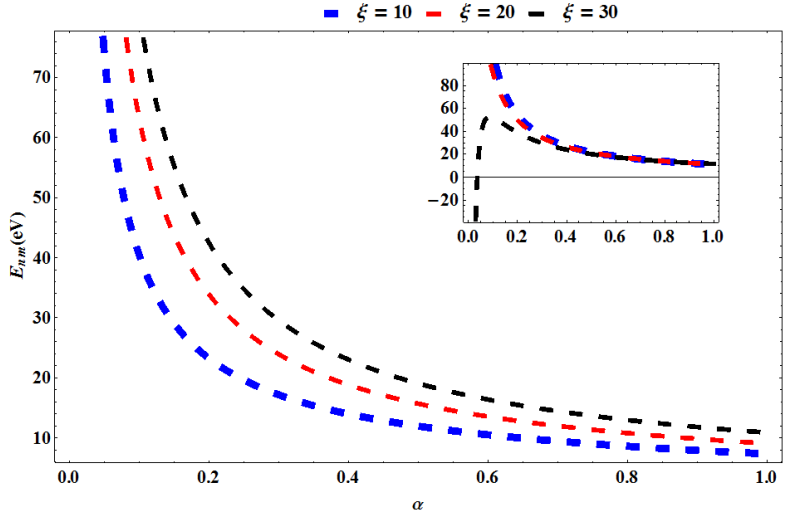

**Figure. 3**: Variation of energy values for the $CO$ diatomic molecule in SMKP under the influence of the magnetic field and AB flux fields with topological defect. Using the fitting parameters $m=n=0$ and $|\vec{B}|=10T$ as a function of $\alpha$ with various $\xi$; the inset is for $|\vec{B}|=100T$.

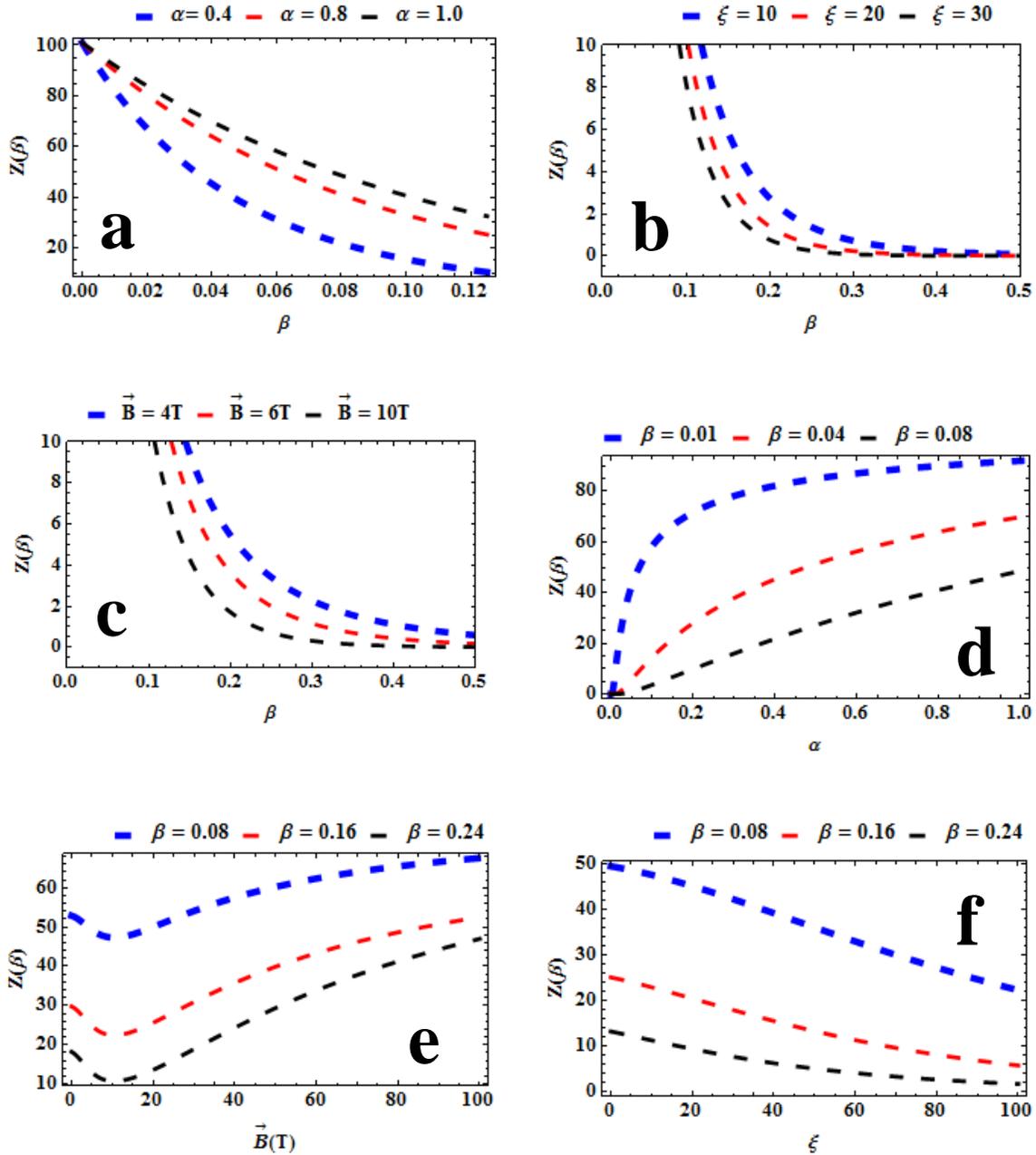

**Figure 4**: Partition function as a function of: **(a)** $\beta$ for different values of $\alpha$. **(b)** as a function of $\beta$ for different values of $\xi$. **(c)** as a function of $\beta$ for different values of $|\vec{B}|$. **(d)** as a function of $\alpha$ different values of $\beta$. **(e)** as a function of $|\vec{B}|$ different values of $\beta$. **(f)** as a function of $\xi$ different values of $\beta$.

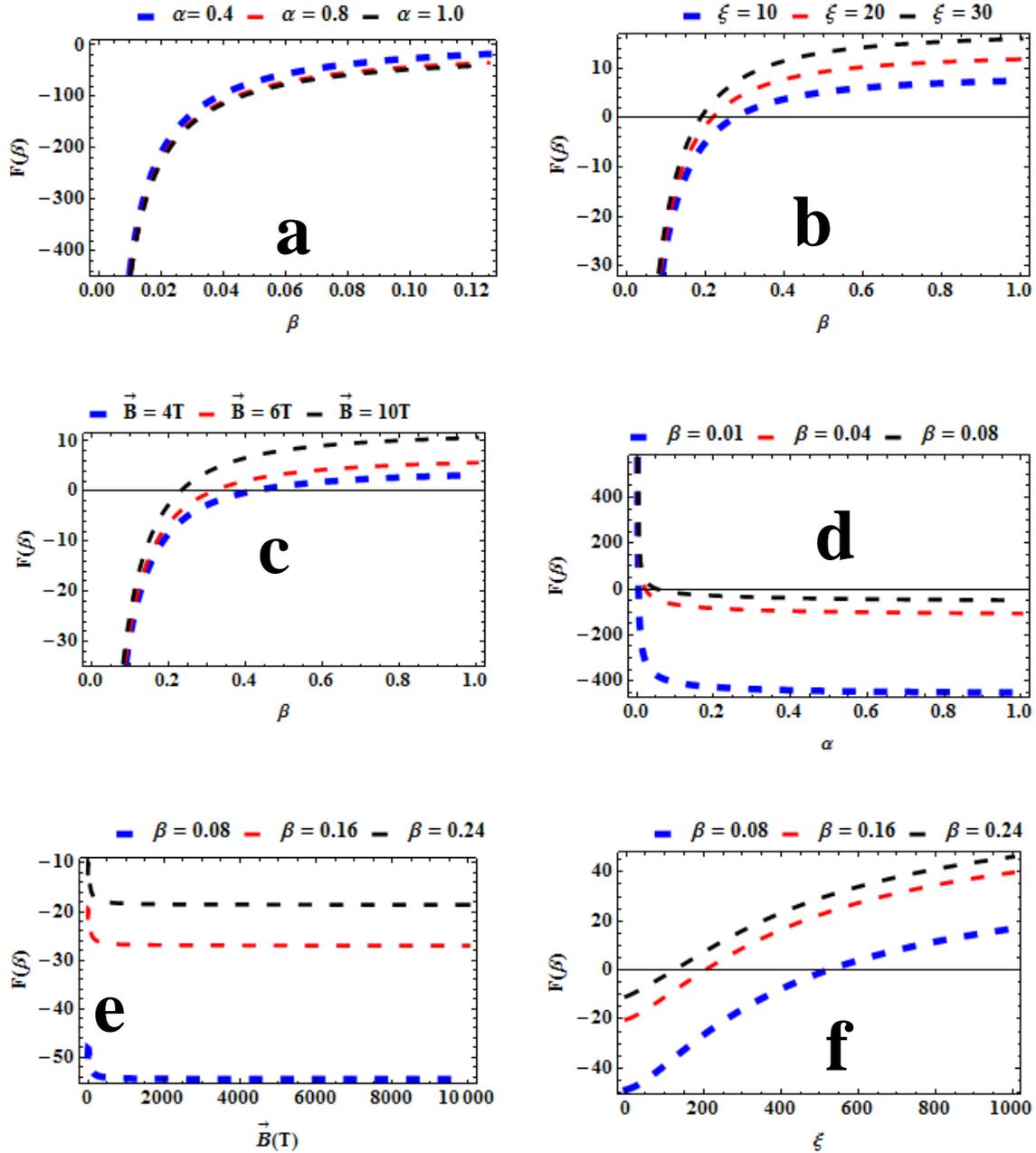

**Figure 5**: Helmholtz Free Energy as a function of: **(a)** $\beta$ for different values of $\alpha$. **(b)** as a function of $\beta$ for different values of $\xi$. **(c)** as a function of $\beta$ for different values of $|\vec{B}|$. **(d)** as a function of $\alpha$ different values of $\beta$. **(e)** as a function of $|\vec{B}|$ different values of $\beta$. **(f)** as a function of $\xi$ different values of $\beta$.

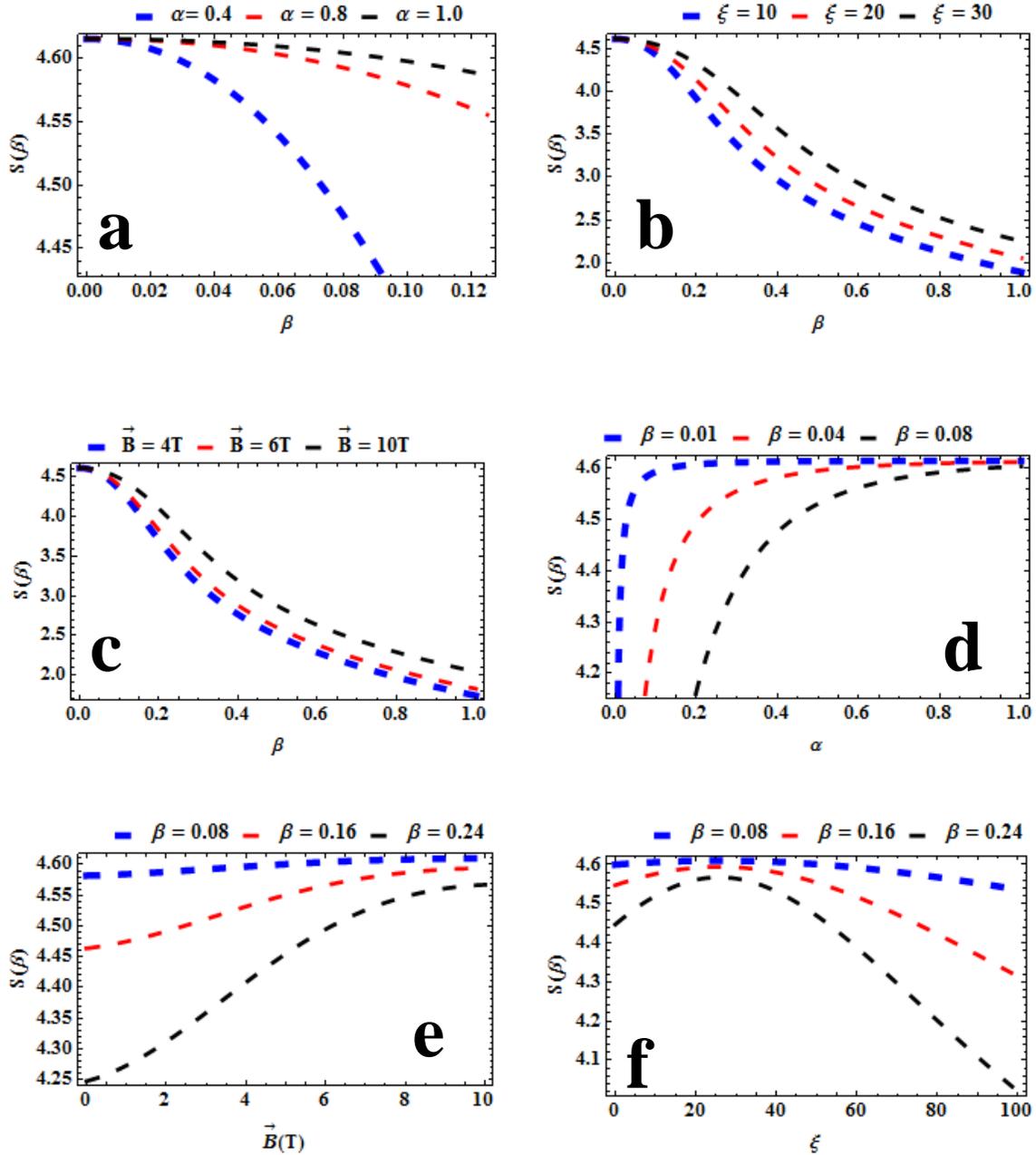

**Figure 6**: Entropy as a function of: **(a)** $\beta$ for different values of $\alpha$. **(b)** as a function of $\beta$ for different values of $\xi$. **(c)** as a function of $\beta$ for different values of $|\vec{B}|$. **(d)** as a function of $\alpha$ different values of $\beta$. **(e)** as a function of $|\vec{B}|$ different values of $\beta$. **(f)** as a function of $\xi$ different values of $\beta$.

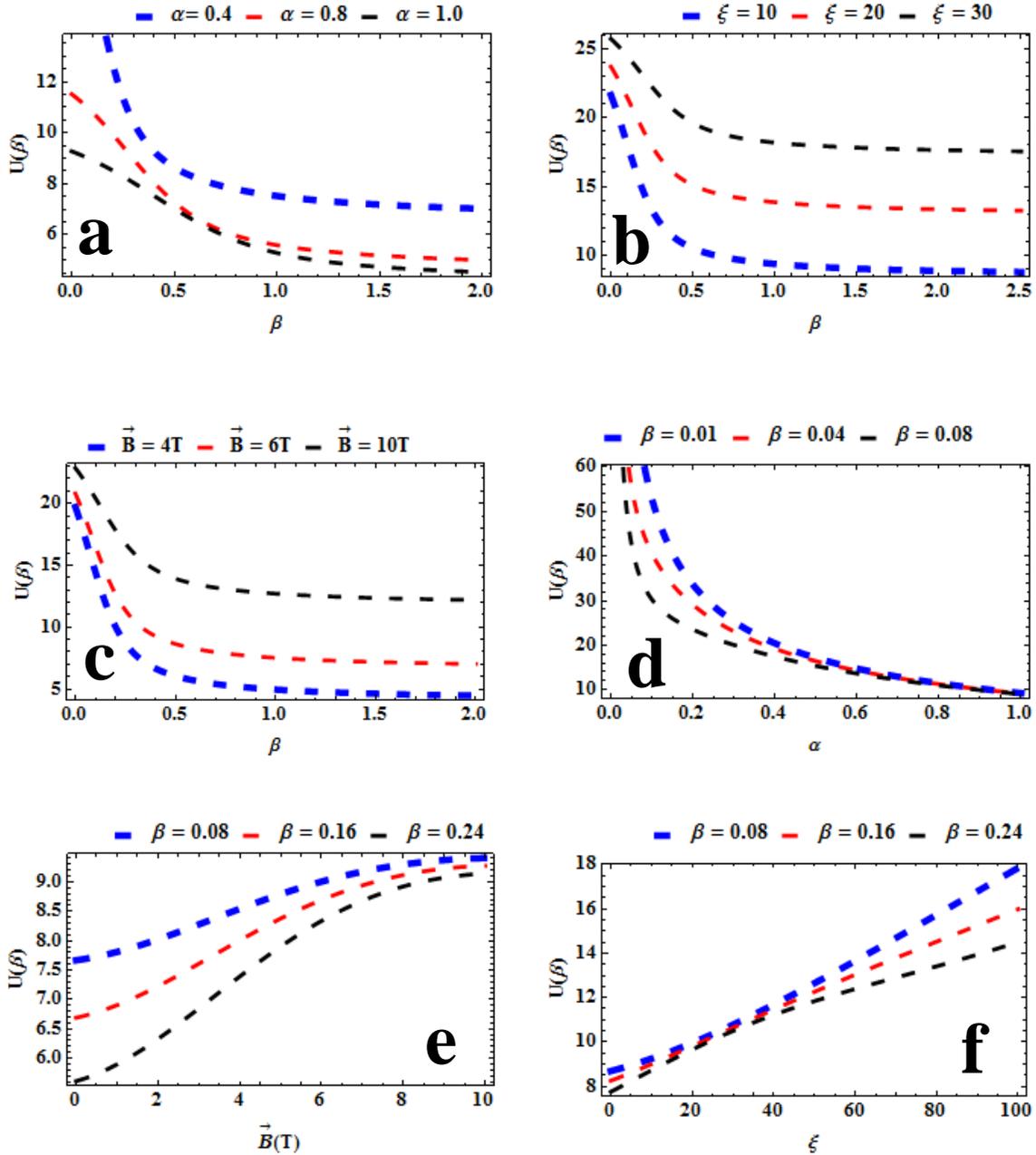

**Figure 7**: Average Energy as a function of: **(a)** $\beta$ for different values of $\alpha$. **(b)** as a function of $\beta$ for different values of $\xi$. **(c)** as a function of $\beta$ for different values of $|\vec{B}|$. **(d)** as a function of $\alpha$ different values of $\beta$. **(e)** as a function of $|\vec{B}|$ different values of $\beta$. **(f)** as a function of $\xi$ different values of $\beta$.

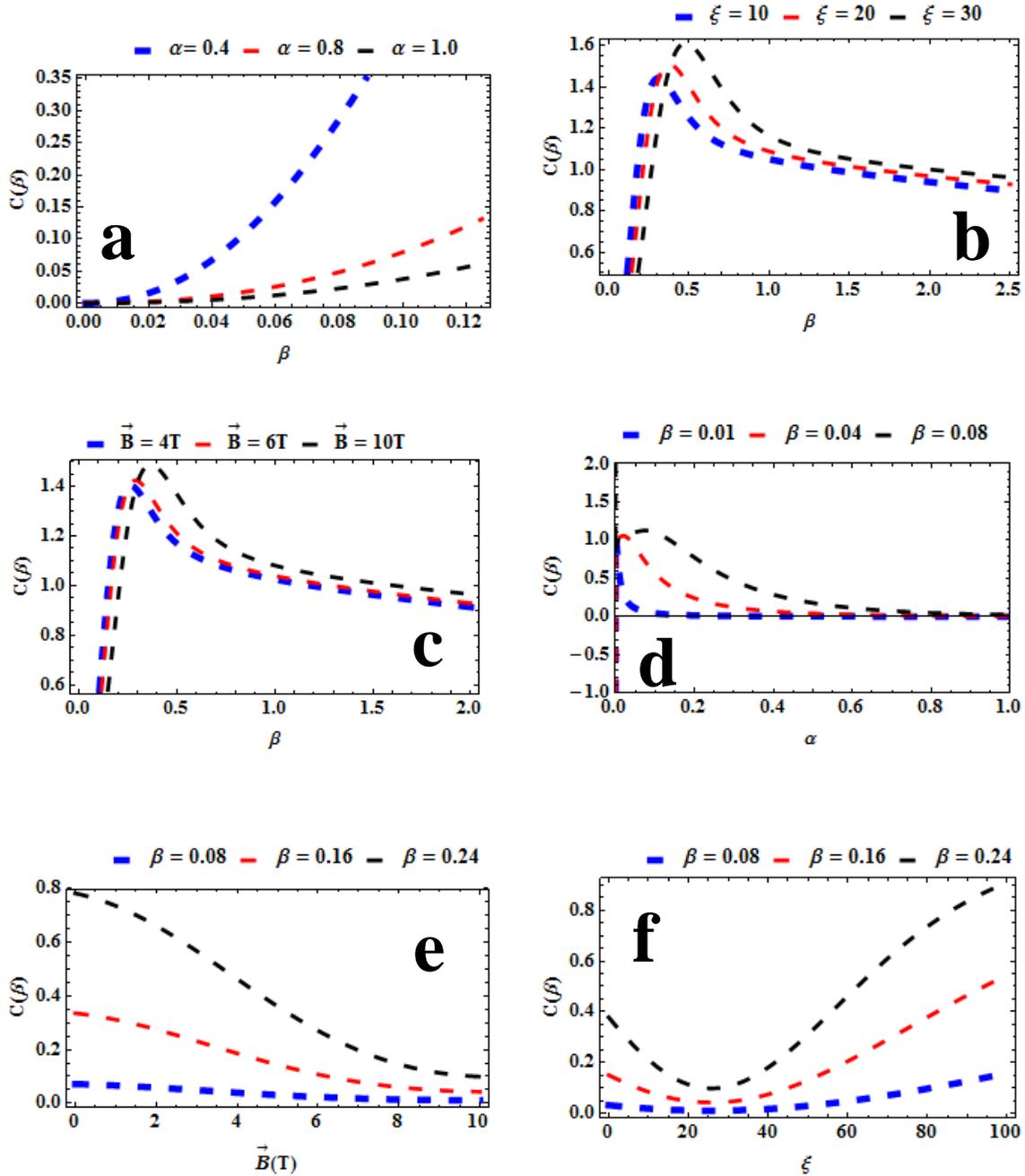

**Figure 8**: Specific Heat Capacity as a function of: **(a)** $\beta$ for different values of $\alpha$. **(b)** as a function of $\beta$ for different values of $\xi$. **(c)** as a function of $\beta$ for different values of $|\vec{B}|$. **(d)** as a function of $\alpha$ different values of $\beta$. **(e)** as a function of $|\vec{B}|$ different values of $\beta$. **(f)** as a function of $\xi$ different values of $\beta$.

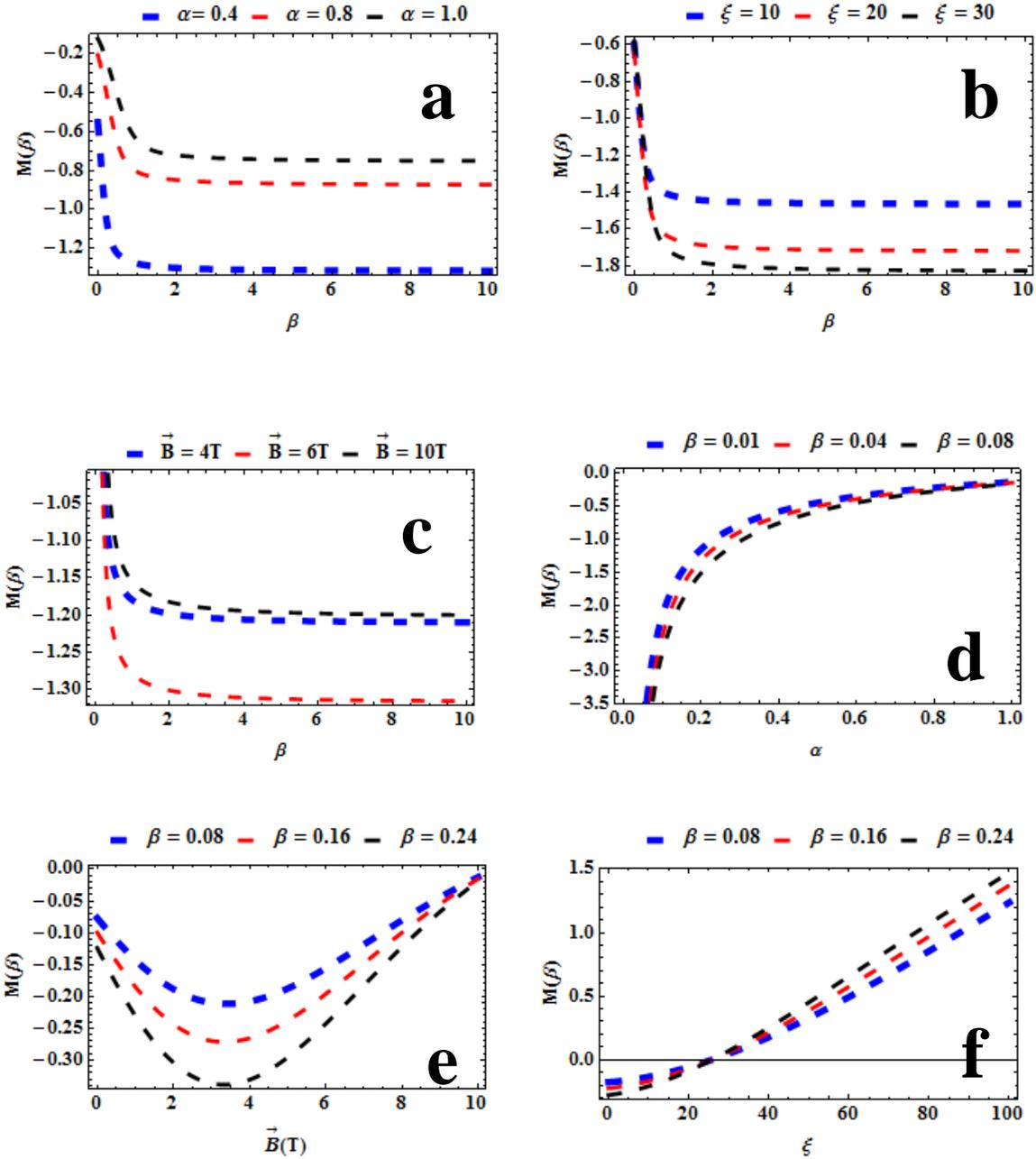

**Figure 9**: Magnetization as a function of: **(a)** $\beta$ for different values of $\alpha$. **(b)** as a function of $\beta$ for different values of $\xi$. **(c)** as a function of $\beta$ for different values of $|\vec{B}|$. **(d)** as a function of $\alpha$ different values of $\beta$. **(e)** as a function of $|\vec{B}|$ different values of $\beta$. **(f)** as a function of $\xi$ different values of $\beta$.

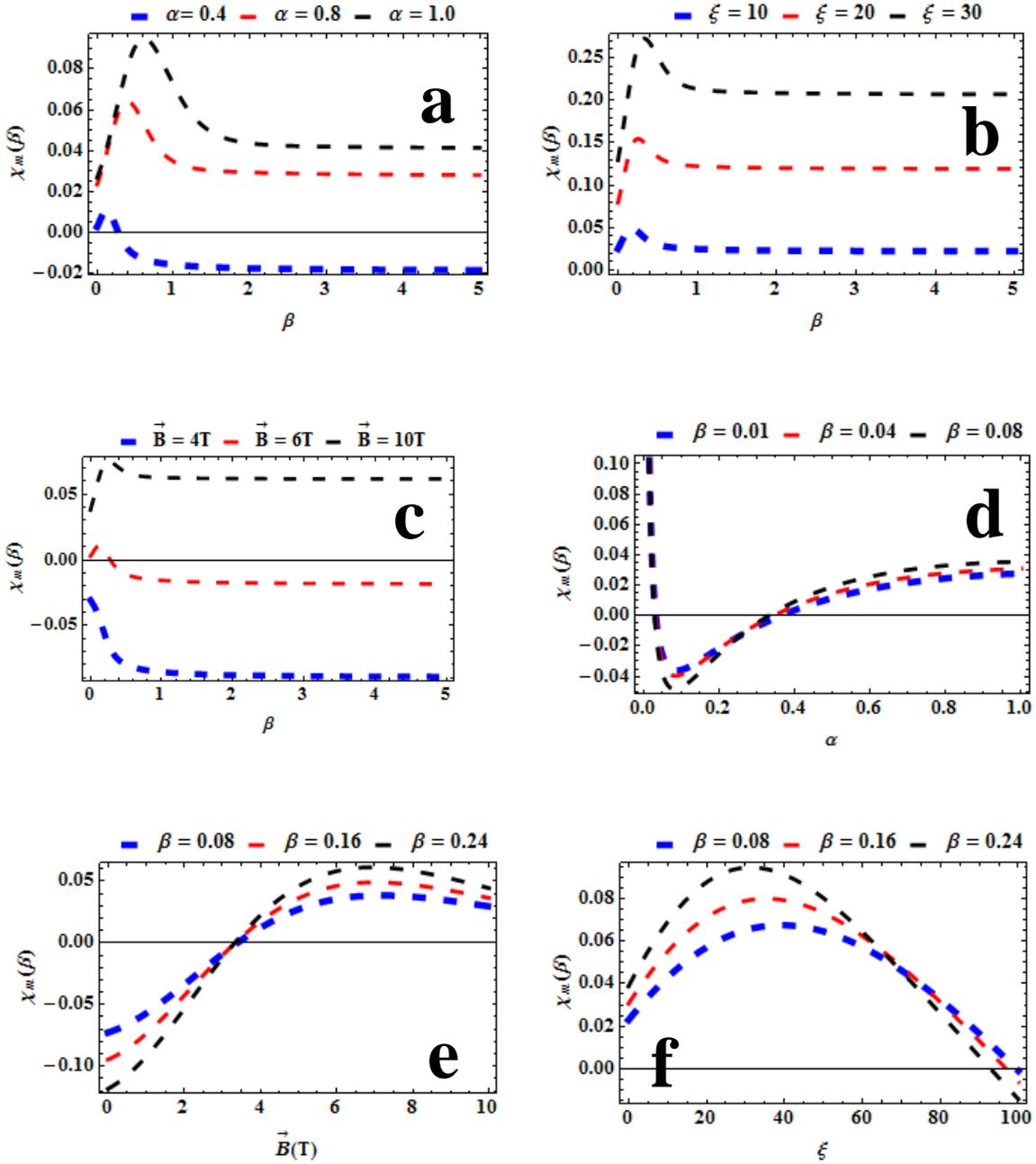

**Figure 10**: Magnetic Susceptibility as a function of: **(a)** $\beta$ for different values of $\alpha$. **(b)** as a function of $\beta$ for different values of $\xi$. **(c)** as a function of $\beta$ for different values of $|\vec{B}|$. **(d)** as a function of $\alpha$ different values of $\beta$. **(e)** as a function of $|\vec{B}|$ different values of $\beta$. **(f)** as a function of $\xi$ different values of $\beta$.

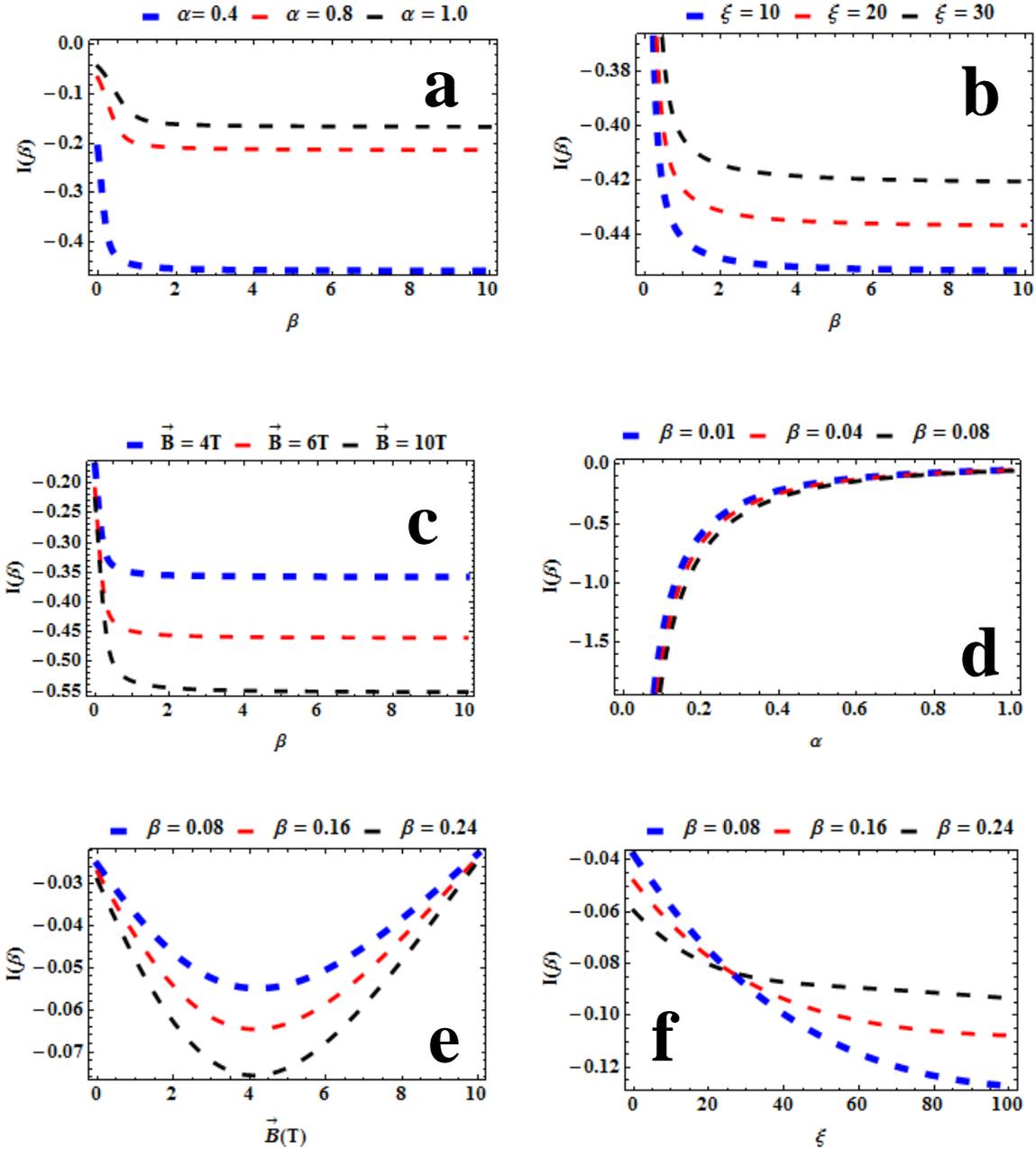

**Figure 11**: Persistent Current as a function of: **(a)** $\beta$ for different values of $\alpha$. **(b)** as a function of $\beta$ for different values of $\xi$. **(c)** as a function of $\beta$ for different values of $|\vec{B}|$. **(d)** as a function of $\alpha$ different values of $\beta$. **(e)** as a function of $|\vec{B}|$ different values of $\beta$. **(f)** as a function of $\xi$ different values of $\beta$.

**Table 1**: The energy levels of the Screened Modified Kratzer potential(SMKP) for $CO$ diatomic molecule with various $n$ and $m$ quantum states for $D=2$ in the presence and absence of external magnetic field, $\vec{B}$ and AB flux field, $\xi$ and topological defect, $\alpha$.

**N/B**: $\alpha = 1$ means the absence topological defect.

| m | n | $\|\vec{B}\|=\xi=0, \alpha=1$ | $\|\vec{B}\|=6T, \xi=0, \alpha=1$ | $\|\vec{B}\|=0, \xi=6, \alpha=1$ | $\|\vec{B}\|=\xi=0, \alpha=0.6$ |
|---|---|---|---|---|---|
| 0 | 0 | 0.1815300 | 3.1621700 | 0.2921620 | 0.2348200 |
|   | 1 | 0.5394900 | 3.4425800 | 0.6489500 | 0.6993420 |
|   | 2 | 0.8892760 | 3.7164800 | 0.9975830 | 1.1556400 |
|   | 3 | 1.2310200 | 3.9839500 | 1.3382000 | 1.6038300 |
| 1 | 0 | 0.1846060 | 3.3134000 | 0.3320690 | 0.2433730 |
|   | 1 | 0.5425330 | 3.5903500 | 0.6884350 | 0.7078250 |
|   | 2 | 0.8922870 | 3.8608200 | 1.0366500 | 1.1640600 |
|   | 3 | 1.2340000 | 4.1249100 | 1.3768600 | 1.6121800 |
| -1 | 0 | 0.1846060 | 3.0140200 | 0.2583760 | 0.2433730 |
|   | 1 | 0.5425330 | 3.2978900 | 0.6155220 | 0.7078250 |
|   | 2 | 0.8922870 | 3.5752000 | 0.9645070 | 1.1640600 |
|   | 3 | 1.2340000 | 3.8460300 | 1.3054700 | 1.6121800 |

| m | n | $\|\vec{B}\|=6T, \xi=6, \alpha=1$ | $\|\vec{B}\|=6T, \xi=0, \alpha=0.6$ | $\|\vec{B}\|=0, \xi=6, \alpha=0.6$ | $\|\vec{B}\|=6T, \xi=6, \alpha=0.6$ |
|---|---|---|---|---|---|
| 0 | 0 | 4.113990 | 3.598980 | 0.345619 | 5.3530800 |
|   | 1 | 4.373410 | 3.995630 | 0.809231 | 5.7191800 |
|   | 2 | 4.626590 | 4.385210 | 1.264630 | 6.0785500 |
|   | 3 | 4.873630 | 4.767810 | 1.711930 | 6.4312800 |
| 1 | 0 | 4.282630 | 4.082520 | 0.415669 | 5.8464800 |
|   | 1 | 4.538510 | 4.470630 | 0.878705 | 6.2041900 |
|   | 2 | 4.788200 | 4.851760 | 1.333540 | 6.5552700 |
|   | 3 | 5.031800 | 5.226010 | 1.780280 | 6.8998000 |
| -1 | 0 | 3.948130 | 3.118530 | 0.292626 | 4.8622300 |
|   | 1 | 4.211070 | 3.523770 | 0.756673 | 5.2367500 |
|   | 2 | 4.467730 | 3.921830 | 1.212510 | 5.6044600 |
|   | 3 | 4.718200 | 4.312820 | 1.660230 | 5.9654300 |

**Table 2**: The bound state energies $E_{n\ell}(eV)$ for $CO$ molecules for different values of the vibrational $n$ and rotational $\ell$ quantum numbers for $(q = 0.4, 0.7 \text{ and } 1.0)$ of the screened Modified Kratzer potential

| | | $q=0.4$ | | $q=0.7$ | | $q=1$ | |
|---|---|---|---|---|---|---|---|
| $n$ | $\ell$ | $CO$(Present) | $CO$(NUFA[11]) | $CO$(Present) | $CO$(NUFA[11]) | $CO$(Present) | $CO$(NUFA[11]) |
| 0 | 0 | 0.112018 | 0.112018 | 0.207914 | 0.207914 | 0.312960 | 0.312960 |
| 1 | 0 | 0.323834 | 0.323834 | 0.609290 | 0.609290 | 0.921983 | 0.921983 |
|   | 1 | 0.324791 | 0.324791 | 0.611132 | 0.611132 | 0.924794 | 0.924794 |
| 2 | 0 | 0.519909 | 0.519909 | 0.992292 | 0.992292 | 1.509746 | 1.509746 |
|   | 1 | 0.520793 | 0.520793 | 0.994048 | 0.994048 | 1.512458 | 1.512458 |
|   | 2 | 0.522561 | 0.522561 | 0.997559 | 0.997559 | 1.517879 | 1.517879 |
| 3 | 0 | 0.700475 | 0.700475 | 1.357198 | 1.357198 | 2.076582 | 2.076582 |
|   | 1 | 0.701287 | 0.701287 | 1.358870 | 1.358870 | 2.079195 | 2.079195 |
|   | 2 | 0.702910 | 0.702910 | 1.362212 | 1.362212 | 2.084422 | 2.084422 |
|   | 3 | 0.705342 | 0.705342 | 1.367223 | 1.367223 | 2.092257 | 2.092257 |
| 4 | 0 | 0.865757 | 0.865757 | 1.704282 | 1.704282 | 2.622814 | 2.622814 |
|   | 1 | 0.866498 | 0.866498 | 1.705871 | 1.705871 | 2.625332 | 2.625332 |
|   | 2 | 0.867978 | 0.867978 | 1.709047 | 1.709047 | 2.630366 | 2.630366 |
|   | 3 | 0.870196 | 0.870196 | 1.713809 | 1.713809 | 2.637914 | 2.637914 |
|   | 4 | 0.873149 | 0.873149 | 1.720151 | 1.720151 | 2.647970 | 2.647970 |
| 5 | 0 | 1.015976 | 1.015976 | 2.033811 | 2.033811 | 3.148763 | 3.148763 |
|   | 1 | 1.016647 | 1.016647 | 2.035318 | 2.035318 | 3.151186 | 3.151186 |
|   | 2 | 1.017987 | 1.017987 | 2.038331 | 2.038331 | 3.156031 | 3.156031 |
|   | 3 | 1.019994 | 1.019994 | 2.042847 | 2.042847 | 3.163295 | 3.163295 |
|   | 4 | 1.022665 | 1.022665 | 2.048862 | 2.048862 | 3.172973 | 3.172973 |
|   | 5 | 1.025997 | 1.025997 | 2.056372 | 2.056372 | 3.185060 | 3.185060 |